\documentclass[a4paper]{article}
\RequirePackage[english]{babel}
\RequirePackage[latin1]{inputenc}
\RequirePackage[T1]{fontenc}
\RequirePackage{mathrsfs}
\RequirePackage{amsmath}
\RequirePackage{amssymb}
\RequirePackage{amsbsy}
\RequirePackage{bm}
\begin{document}
\title{\bf{Massless Fermion Mixing for\\ Semispinorial Torsional Interaction}}
\author{Luca Fabbri\\
\footnotesize{INFN, Sec. Bologna \& Dep. Physics, Univ. Bologna, 40126 Bologna, Italy}}
\date{}
\maketitle
\begin{abstract}
We consider a geometric approach to field theory in which torsion is present in the matter field equations, and we develop the consequences of the torsion-spin coupling for a pair of single-handed spinors; we show these interactions to have the structure of a flavour-changing neutral current giving rise to a mixing for two massless fermion fields: as torsional interactions for semispinors entail mixing for massless fermions we speculate that this mechanism could be responsible for oscillations of neutrinos.
\end{abstract}
\section*{Introduction}
Neutrinos have first been introduced by Pauli, who defined them to be fermions with no mass nor charge, so that they can carry energy while being undetectable, explaining the problem of missing energy in the $\beta$-decay; further evidence eventually established that neutrinos come in three types. The fact that neutrinos of these three flavours could change into each other was then theorized by Pontecorvo, who built a model for which neutrinos could mix into one another, so that one of them could be converted into any one of the two others, explaining the fact that we observe only one third of all neutrinos produced in the process of the $\beta$-decay having place in the sun (\cite{p/1}, \cite{p/2} and \cite{g-p}); on the other hand, for such neutrino oscillation to happen, these particles were assumed to be massive and with masses of different values. After the model of neutrino oscillations for massive non-degenerate states has been proposed, several generalizations have been pursued: at first by postulating that the flat spacetime is curved by gravitation (see \cite{a-b}, \cite{c-f} and \cite{p-r-w}), and then by assuming that the curved spacetime is twisted by torsion (see \cite{a-s} and \cite{a-d-r}); an interesting model was propounded, in which neutrino oscillations for massless although always double-handed states were discussed, again in twisted-curved spacetime (see \cite{s-g}). In these generalizations, if not the mass at least the double-handedness must be present, and this has two issues: one is the obvious fact that particles with masses of different values, if produced with the same energy, have different speeds, and so even if the packets were produced at the same location, they would soon be separated making superposition fail (\cite{a-s}, \cite{a-d-r}); the other issue is that even in the case of massless spinors, there is anyway double-handedness requiring one projection to interact while the other projection is sterile (\cite{s-g}). These issues are problematic because they are related to the observational fact that never have these masses been observed, neither is justified to assume these fields to be massless but to postulate that they still have two physical components nor is satisfactory to have that of these two physical components one cannot be detected. To adhere to the observational facts, the only justified and satisfactory solution appears to be constituted by single-handed states. As a matter of fact, in the literature there have also been models in which single-handed neutrinos had specific interactions with off-diagonal terms able to induce oscillations, and therefore massless neutrinos could have interactions with other fermions while propagating throughout matter (see \cite{m-s} and \cite{w}). However in this case too there are two different types of problems: one was that such oscillations were due to interactions whose off-diagonal terms are not part of the standard model of forces, the other is that even if such interactions were actually present the oscillations they induce would be impossible in the material vacuum. Of course one may now wonder what would happen if these models were to be allowed to have torsion, but as our knowledge is concerned no such investigation has ever been performed. To conciliate both approaches, those having no additional interactions but double-handedness (\cite{s-g}) and those having single-handedness but with additional interactions (\cite{w}), we should have no additional interactions and single-handedness, and torsion for semispinor may therefore be the solution.

In the following, we shall consider the specific example represented by the spinor in massless state, which will be taken to be spinors with single-handed structure: we will see that because of the presence of torsion in the covariant derivatives, a torsion-spin interaction arises for the dynamics of the semispinors, and we are going to see how these torsional interactions among semispinors produce a mixing for massless fermion fields.
\section{Fermion Mixing from Spinorial Torsion}
As we have stated above, our starting point will be to consider the field equation for semispinors 
\begin{eqnarray}
&i\gamma^{\mu}D_{\mu}\nu_{1}=0\\
&i\gamma^{\mu}D_{\mu}\nu_{2}=0
\end{eqnarray}
in terms of the most general covariant derivatives $D_{\mu}$ that can be decomposed into the simplest covariant derivatives $\nabla_{\mu}$ as in the torsionless case plus torsional contributions as
\begin{eqnarray}
&i\gamma^{\mu}\nabla_{\mu}\nu_{1}
+\frac{3}{16}\bar{\nu}_{2}\gamma_{\mu}\nu_{2}\gamma^{\mu}\nu_{1}=0\\
&i\gamma^{\mu}\nabla_{\mu}\nu_{2}
+\frac{3}{16}\bar{\nu}_{1}\gamma_{\mu}\nu_{1}\gamma^{\mu}\nu_{2}=0
\end{eqnarray}
with spinorial interactions of each semispinor with the other; these have now to be written in an alternative form by employing Fierz rearrangements so to get the equations
\begin{eqnarray}
&i\gamma^{\mu}\nabla_{\mu}\nu_{1}
-\frac{1}{16}\left(\bar{\nu}_{1}\gamma_{\mu}\nu_{1}
-\bar{\nu}_{2}\gamma_{\mu}\nu_{2}\right)\gamma^{\mu}\nu_{1}
-\frac{1}{8}\left(\bar{\nu}_{2}\gamma^{\mu}\nu_{1}\right)\gamma_{\mu}\nu_{2}=0\\
&i\gamma^{\mu}\nabla_{\mu}\nu_{2}
-\frac{1}{8}\left(\bar{\nu}_{1}\gamma^{\mu}\nu_{2}\right)\gamma_{\mu}\nu_{1}
+\frac{1}{16}\left(\bar{\nu}_{1}\gamma_{\mu}\nu_{1}
-\bar{\nu}_{2}\gamma_{\mu}\nu_{2}\right)\gamma^{\mu}\nu_{2}=0
\label{fieldequations}
\end{eqnarray}
in which the interactions of each semispinor with the other has been written in a form that is particularly interesting; in fact, such a coupled system of field equations can be written as
\begin{eqnarray}
&\!\!\!\!i\gamma^{\mu}\!
\left[\nabla_{\mu}\!\!
\left(\!\!\!\!\begin{tabular}{c}$\nu_{1}$\\$\nu_{2}$\end{tabular}\!\!\!\!\right)
\!+\!\frac{i}{16}\!\left(\!\!\!\!\begin{tabular}{cc}
$(\bar{\nu}_{1}\gamma_{\mu}\nu_{1}-\bar{\nu}_{2}\gamma_{\mu}\nu_{2})$ & $2(\bar{\nu}_{1}\gamma_{\mu}\nu_{2})^{\ast}$\\ $2(\bar{\nu}_{1}\gamma_{\mu}\nu_{2})$ & $-(\bar{\nu}_{1}\gamma_{\mu}\nu_{1}-\bar{\nu}_{2}\gamma_{\mu}\nu_{2})$ \end{tabular}\!\!\!\!\right)\!\!
\left(\!\!\!\!\begin{tabular}{c}$\nu_{1}$\\ $\nu_{2}$\end{tabular}\!\!\!\!\right)\!\right]\!=\!0
\end{eqnarray}
in which we achieve a simplification as
\begin{eqnarray}
&i\gamma^{\mu}\left[\nabla_{\mu}\nu+ig\vec{A}_{\mu}\cdot\frac{\vec{\sigma}}{2}\nu\right]=0
\label{fieldequationssimplified}
\end{eqnarray}
when the vectors $\vec{A}_{\mu}$ are introduced; a last step to accomplish is the final compactification given by 
\begin{eqnarray}
&i\gamma^{\mu}\mathbb{D}_{\mu}\nu=0
\label{fieldequationssimplifiedcompactified}
\end{eqnarray}
once the vectors $\vec{A}_{\mu}$ and the free covariant derivative $\nabla_{\mu}$ are together written as the interacting covariant derivative $\mathbb{D}_{\mu}$ when this is defined. What permitted us to write the field equations (\ref{fieldequationssimplified}) was the definition of the doublet of spinors
\begin{eqnarray}
&\left(\!\!\begin{tabular}{c}$\nu_{1}$\\$\nu_{2}$\end{tabular}\!\!\right)=\nu
\label{doublet}
\end{eqnarray}
with which to define the triplet of vectors
\begin{eqnarray}
&\frac{1}{8}\bar{\nu}\gamma_{\mu}\vec{\sigma}\nu=g\vec{A}_{\mu}
\label{triplet}
\end{eqnarray}
in terms of the constant $g$ that has been included for generality; the passage from the field equation with the form given in (\ref{fieldequationssimplified}) to the field equations with form given in (\ref{fieldequationssimplifiedcompactified}) was possible after defining
\begin{eqnarray}
&\mathbb{D}_{\mu}\nu=\nabla_{\mu}\nu+ig\vec{A}_{\mu}\cdot\frac{\vec{\sigma}}{2}\nu
\label{derivative}
\end{eqnarray}
where the triplet of vectors $\vec{A}_{\mu}$ and the triplet of rank-$2$ matrices that generate the infinitesimal transformation of the $SU(2)$ group are combined into the gauge connection of the $SU(2)$ group, and eventually these are added to the free covariant derivative $\nabla_{\mu}$ to give the gauge covariant derivative $\mathbb{D}_{\mu}$ associated to the $SU(2)$ group. We remark that the covariance under $SU(2)$ transformations is local because the semispinors are functions of the spacetime position and so their $SU(2)$ transformations may take place with coefficients that depend on the spacetime position themselves, and so for $SU(2)$ transformations with local parameters we have that the transformation of the doublet of spinors induces the transformation on the triplet of vectors given by 
\begin{eqnarray}
&\nu'=e^{ig\vec{\theta}\cdot\frac{\vec{\sigma}}{2}}\nu\\
&\left[\vec{A}_{\mu}\cdot\frac{\vec{\sigma}}{2}\right]'
=e^{ig\vec{\theta}\cdot\frac{\vec{\sigma}}{2}}
\left[\left(\vec{A}_{\mu}-\partial_{\mu}\vec{\theta}\right)\cdot\frac{\vec{\sigma}}{2}\right]
e^{-ig\vec{\theta}\cdot\frac{\vec{\sigma}}{2}}
\label{transformation}
\end{eqnarray}
once (\ref{triplet}) are considered; gauge covariant derivatives transform as
\begin{eqnarray}
&\left(\mathbb{D}_{\mu}\nu\right)'=e^{ig\vec{\theta}\cdot\frac{\vec{\sigma}}{2}}\mathbb{D}_{\mu}\nu
\label{derivativetransformation}
\end{eqnarray}
once the definition (\ref{derivative}) is given. These field equations are invariant for $SU(2)$ local transformations because of (\ref{derivativetransformation}) as it is obvious.

Now, the evolution of the couple of fields is governed by the evolution equations defined above as
\begin{eqnarray}
&\nu'=e^{ig\vec{\theta}\cdot\frac{\vec{\sigma}}{2}}\nu\\
&\left[\vec{A}_{\mu}\cdot\frac{\vec{\sigma}}{2}\right]'
=e^{ig\vec{\theta}\cdot\frac{\vec{\sigma}}{2}}
\left[\left(\vec{A}_{\mu}-\partial_{\mu}\vec{\theta}\right)\cdot\frac{\vec{\sigma}}{2}\right]
e^{-ig\vec{\theta}\cdot\frac{\vec{\sigma}}{2}}
\end{eqnarray}
which may be expanded in terms of infinitesimal parameters, giving
\begin{eqnarray}
&\nu'=\nu+\frac{i}{2}g\delta\vec{\theta}\cdot\vec{\sigma}\nu\\
&\vec{A}_{\mu}'=\vec{A}_{\mu}
-\partial_{\mu}\delta\vec{\theta}+g\vec{A}_{\mu}\times\delta\vec{\theta}
\end{eqnarray}
which we are now going to study separately by focusing on the spinorial evolution and treating the gauge fields as external interactions that may be considered to be constant in this approximation, so that
\begin{eqnarray}
&\nu'=\nu+\frac{i}{2}g\delta\vec{\theta}\cdot\vec{\sigma}\nu\\
&\partial_{\mu}\delta\vec{\theta}=g\vec{A}_{\mu}\times\delta\vec{\theta}
\end{eqnarray}
telling us that Pauli matrices with non-diagonal structure appear as soon as the infinitesimal parameters are different from zero and that such infinitesimal parameters must be different from zero at some time of their evolution. To see this, we consider the circumstance in which in the initial configuration there is only one spinor field $\nu_{1}$ so that correspondingly there is only one gauge vector given by $A_{\mu}^{3}$ and therefore
\begin{eqnarray}
&\nu'_{1}=\nu_{1}+\frac{i}{2}g\delta\theta^{3}\nu_{1}\\
&\nu'_{2}=\frac{i}{2}g(\delta\theta^{1}+i\delta\theta^{2})\nu_{1}
\end{eqnarray}
with
\begin{eqnarray}
&\partial_{\mu}\delta\theta^{1}=-g\delta\theta^{2}A_{\mu}^{3}\\
&\partial_{\mu}\delta\theta^{2}=g\delta\theta^{1}A_{\mu}^{3}\\
&\partial_{\mu}\delta\theta^{3}=0
\end{eqnarray}
proving that $\nu_{2}$ appears if one of the infinitesimal parameters $\delta\theta^{1}$ or $\delta\theta^{2}$ does not vanish and that the two parameters $\delta\theta^{1}$ and $\delta\theta^{2}$ do not vanish at later times since their evolution is given in terms of circular functions as
\begin{eqnarray}
&\nu'_{1}=\nu_{1}+\frac{i}{2}g\delta\theta^{3}\nu_{1}\\
&\nu'_{2}=\frac{i}{2}g(\delta\theta^{1}+i\delta\theta^{2})\nu_{1}
\end{eqnarray}
with
\begin{eqnarray}
&\delta\theta^{1}=a\cos{\left(gA_{\mu}^{3}x^{\mu}\right)}-b\sin{\left(gA_{\mu}^{3}x^{\mu}\right)}\\
&\delta\theta^{2}=b\cos{\left(gA_{\mu}^{3}x^{\mu}\right)}+a\sin{\left(gA_{\mu}^{3}x^{\mu}\right)}\\
&\delta\theta^{3}=k
\end{eqnarray}
as it can be checked by directly performing all calculations. In this way by choosing the integration constants to be $k=0$ with $a=0$ and $b=-2$ we finally have the following
\begin{eqnarray}
&\nu'_{1}=\nu_{1}\\
&\nu'_{2}=ge^{i\left(gA_{\mu}^{3}x^{\mu}\right)}\nu_{1}
\end{eqnarray}
showing that the initial single semispinor is later accompanied by a second semispinor. Therefore, the presence of torsion for semispinors implies that it is possible to have a mixing of massless fermion fields according to $SU(2)$ gauge transformation; under certain approximations even if we start from a single semispinor we will obtain another semispinor appearing later. Clearly if we want a precise treatment of this problem then we have to renounce to such approximations dealing with the full non-linear field equations, although for such a study numerical simulations will be necessary and therefore it is beyond the scope of the present discussion.
\section*{Conclusion}
In this paper, we have seen that the presence of torsion for semispinors implies the mixing for massless fermion fields having the $SU(2)$ structure; we have shown that under certain approximations even if we start with a configuration having only one semispinor there is a second semispinor appearing later. If we want to have not only a qualitative behaviour but also quantitative results then the full non-linear system of equations has to be solved numerically. 

The interpretation we could give to describe this situation is that the torsionally propagating semispinors feel the torsion-spin interactions through which they combine into vectors that are carriers of the exchanged flavour for the mixing of massless fermion fields, that is torsion forces semispinors to have a flavour-changing neutral current determining the mixing of the massless fermions; notice that torsionally induced flavour-changing neutral currents are present precisely because the semispinors are massless.

It is tempting to speculate that such mechanism can be the explanation of oscillations of neutrinos.

\end{document}